\documentclass[12pt]{article}
\usepackage{graphicx}
\usepackage{dcolumn}
\usepackage{bm}
\usepackage{graphics}
\usepackage{amsmath}
\usepackage{amssymb}
\usepackage{amscd}
\usepackage{afterpage}
\usepackage{float,times}
\usepackage{subfigure}
\usepackage{rotating}
\usepackage{multirow}
\usepackage{epsfig}
\usepackage{theorem}
\usepackage{moreverb}
\usepackage{euscript}
\usepackage{psfrag}

\begin{document}

\begin{titlepage}
\begin{center}
%{\hbox to\hsize{
%\hfill \bf hep-th/??? }}
{\hbox to\hsize{\hfill June 2008 }}

\bigskip
\vspace{6\baselineskip}

{\Large \bf

CP violation and gravity as the weakest force \\
}
\bigskip

\bigskip

{\bf 
~~
 Archil Kobakhidze 
 \\}
\smallskip

{ \small \it 
School of Physics, The University of Melbourne, Victoria 3010, Australia \\ 
E-mail: archilk@unimelb.edu.au \\ }

\bigskip

\vspace*{.5cm}

{\bf Abstract}\\
\end{center}
\noindent
{\small 
We argue that CP violation has rather dramatic impact on the "gravity as the weakest force" conjecture. Namely we find that new ultraviolet scale must be $\Lambda \lesssim \theta g^3 M_P$, where $\theta$ is an effective parameter describing CP violation and $g$ is the gauge coupling constant. 
The bound implies that CP-conserving limit is discontinuous, and possibly indicates that the class of effective theories with strict CP conservation is inconsistent with a fundamental theory incorporating quantum gravity. At the same time, the mass hierarchy problem can be explained due to the smallness of the CP-violation, $\theta \sim 10^{-15}$ or so. } 
\bigskip
\bigskip
\bigskip
\end{titlepage}
\baselineskip=16pt
%\paragraph{1.}
%\paragraph{Introduction} 
The mass hierarchy problem can be formulated as the question: why is the gravity so weak compared to other observed fundamental interactions? Simple perturbative arguments suggest that the masses of weak bosons, for example, emerging upon the spontaneous symmetry due to the Higgs mechanism, must be of the order of Planck mass $M_P\approx 10^{18}$ GeV. Consequently the weak interactions must have the gravitational strength. Instead we observe that the weak bosons are some 16 orders of magnitude lighter, and the strength of weak interactions is $\approx 10^{32}$ times stronger than the strength of gravitational interactions. One way to think about this puzzle is that the perturbative arguments somehow break down at energy scales of the order of a few TeV. That is to say, the effective quantum field theory has an intrinsic cut-off $\Lambda \approx $ few TeV. 
Recently, based on some nonperturbative considerations related with black hole physics, it has been argued in \cite{ArkaniHamed:2006dz} and \cite{Dvali:2007hz}-\cite{Dvali:2008tq}, that the actual cut-off of the effective field theory can indeed be smaller than the Planck mass, $\Lambda < M_P$. The crucial role in the argumentation of \cite{ArkaniHamed:2006dz} is played by the conserved local electric and magnetic charges which can be monitored outside the black hole by long-range classical potentials. In \cite{Dvali:2007hz}, \cite{Dvali:2008tq}, global discrete $Z_N$ charges are considered, which emerge as a result of spontaneous symmetry breaking $U(1)\to Z_N$. The black hole in this case have quantum hair associated with discrete $Z_N$ charges which are detectable through the Aharonov-Bohm scattering of $U(1)$ cosmic strings off the black hole. The bound on the cut-off obtained in \cite{Dvali:2008tq} is different from the one discussed in \cite{ArkaniHamed:2006dz}. Also, in \cite{Dvali:2007hz} and \cite{Dvali:2007wp}, instead of $Z_N$, simply large N-number species of particles has been considered, and the bound on the cut-off obtained there is essentially the same as in the $Z_N$ case. If $N\approx 10^{32}$, then, as has been argued in \cite{Dvali:2007hz}-\cite{Dvali:2008tq}, the stable hierarchy between the electroweak and Planck scales can be achieved. 

In this brief note we show that extending the line of reasoning of ref. \cite{ArkaniHamed:2006dz} for CP violating case, one obtains much stronger bound on the field theory cut-off $\Lambda$. The hierarchy then can be attributed to the smallness of the CP violation. The fact that the hierarchy problem can be solved by inducing small electric charge for the monopole is also evident in \cite{Dvali:2007hz}. However, the bound obtained in \cite{Dvali:2007hz} is looser then the one discussed below. 
%\paragraph{CP violation and the bound on the cut-off } 

Let us start by considering the Reissner-Nordstrom charged black hole of mass $M_{\rm BH}$ and a charge $Q$. The charge $Q$ in general comprises of electric $Q_e$ and magnetic $Q_m$ contributions, $Q=\sqrt{Q_e^2+Q_m^2}$. The necessary condition for the existence of such a black hole is $M_{BH}\geq Q M_P$. We must assume that this black hole radiates its charge completely, so that there are no charged remnants (see \cite{Susskind:1995da} for reasoning against stable remnants).  Obviously, particles that carry away the black hole charges must be lighter than the extremal black hole of a given charge. For example, a particle with the minimal electric charge $g_e$ must have the mass, 
\begin{equation}
m_{\rm el}\leq g_eM_{P}~.
\label{1}
\end{equation} 
Analogously, the mass of the magnetically charged particle with the elementary magnetic charge $g_m\sim 1/g_e$ must be bounded, 
\begin{equation}
m_{\rm mon}\leq g_{m}M_{P}\sim \frac{M_P}{g_e}~.
\label{2}
\end{equation}
Recall that the contribution of the magnetic field energy to the mass of the monopole is linearly divergent and must be regularized. Assuming a cut-off $\Lambda$, one obtains, $m_{\rm mon}\sim \frac{\Lambda}{g_e^2}$. Plugging this mass into the eq. (\ref{2}), one obtains the following bound on the cut-off \cite{ArkaniHamed:2006dz}:
\begin{equation}
\Lambda \lesssim g_e M_P~. 
\label{3}
\end{equation} 
In the perturbative case $g_e<1$ the cut-off turns out to be smaller than the Plank mass, contrary to the naive expectation. 
Now we will argue that the above bound is modified significantly in a CP-violating theory. The key difference introduced by CP-violation is that there are no pure magnetic monopoles \cite{Witten:1979ey}, as well as there are nor pure magnetically charged black holes \cite{Kobakhidze:2008em} in CP violating theories. Indeed, if we denote the parameter describing strength of the CP-violation in the effective monopole action by $\theta$, the monopole acquires the minimal electric charge $g_e^{\rm ind}\sim \theta g_e$. With small CP-violation, $\theta < 1$, this induced charge is by no means smaller than $g_e$. As a result, we obtain the bound on the cut-off, 
\begin{equation}
\Lambda \lesssim \theta g_e^3 M_P~.
\label{4}
\end{equation}
This bound is considerably stronger than the bound in (\ref{3}). 

Two important consequences can be immediately deduced from our bound (\ref{4}). First, the limit $\theta \to 0$ is not achievable, as it implies $\Lambda \to 0$. That is to say, CP-violating and CP-conserving sectors of the landscape are disentangled in a sense that $\theta \to 0$ limit 
is discontinuous. It is conceivable to think then that theories describing strictly CP-conserving worlds must represent the "swampland" \cite{Vafa:2005ui}, i.e. the class of effective theories possibly inconsistent with quantum gravity. 
Second, the possible solution to the mass hierarchy problem, according to (\ref{4}), can be attributed to the smallness of the CP-violation, e.g. $\theta \sim 10^{-15}$ or so\footnote{Note that the solution of the hierarchy problem we understand in a restricted sense. The low cut-off actually implies that quantum corrections do not considerably modify tree level masses, resolving the "technical" aspect of the problem.}. Note, that $\theta$ above is viewed as a parameter emerging in the effective action for monopoles as a result of quantum corrections from the CP-violating sector of the theory. Thus, $\theta$ is model dependent parameter, and its explicit calculation is not the easy task. However, in view of smallness of the observed CP violation (e.g. the bound on electric dipole moment of neutron implies the strong CP parameter to be $\lesssim 10^{-10}$) one might think that to have $\theta \sim 10^{-15}$ is not unrealistic. Also, the induced electric charge can straightforwardly emerge from the CP-violating topological term, $\sim \theta F_{\mu\nu}\tilde F^{\mu\nu}$. Since such a term does not renormalize beyond one-loop, we expect the induced charge will be one-loop exact. 

In deriving the bound (\ref{4}) we have relied on the existence of magnetic monopoles which acquire small electric charges if CP invariance is broken. Actually, the similar physical situation can be realized in theories where instead of magnetic monopoles one has massive higher integer spin (e.g. spin-2) quantum hair \cite{Dvali:2006az}. Indeed, formally the massive quantum hair field configuration is the same as the monopole field configuration. The physical difference between the two is that, while the monopole produces a local field, the massive quantum hair is not observable locally. However, if one assumes CP-violating topological coupling between an unbroken U(1) field strength $F_{\mu\nu}$ and the hair providing field $H_{\mu\nu}$ ($H_{ij}=\mu\epsilon_{ijk}x^k/r$) , $\sim \theta F_{\mu\nu}\tilde H^{\mu\nu}$, then particles carrying hidden magnetic charges associated with the quantum hair will acquire induced U(1) electric charges $g_e^{\rm ind}\sim \theta \mu$, just like it happens in the presence of monopoles. Then there must exist light particles with masses, 
\begin{equation}
m_{\rm el}\lesssim \theta \mu M_{P}~.
\label{a1}
\end{equation}
The bound on the cut-off can not be derived in this case, because the quantum hair, being purely topological, does not contribute to the particle mass.
 
There is some close resemblance between the relevance of CP violating topological $\theta$-term discussed above and the role of the similar terms in obtaining the bounds in case of global discrete charges \cite{Dvali:2007hz}- \cite{Dvali:2008tq}. Indeed the quantum hair associated with those global charges can only be probed quantum mechanically if the hair providing field couples to a string through the boundary term \cite{Dvali:2006az}
\begin{equation}
\theta \int dX^{\mu}\wedge dX^{\nu}H_{\mu\nu}~,
\end{equation}
where $X^{\mu}$ are the string target space coordinates. A particle (black hole) of mass $m$ carrying such a quantum hair scatters off the string if $\mu \theta \neq n/2$, $n$ is an integer. That is to say, the particle can have additional unobservable global charge $Q=2\mu \theta$ (n=1), which is analogous of our induced local electric charge $g_{e}^{\rm ind}$. It has been argued further in \cite{Dvali:2007hz} that $Q$ is quantized in units of $1/N$, where $N\lesssim \frac{M_P^2}{m^2}$ ($N$ is the number of particle species). This implies the bound on the mass of a particle, 
\begin{equation}
m^2\lesssim Q M_P^2\approx \theta \mu M_P^2~,
\label{a2}
\end{equation}
derived first in \cite{Dvali:2007hz}. Again we see that the particle mass can be much smaller than the Planck mass, providing the induced charge Q is small. However, the bound (\ref{a2}) is different from the one in (\ref{a1}). It will be interesting to better understand possible relation between the present work and those in \cite{Dvali:2007hz}- \cite{Dvali:2008tq}. 

In conclusion, we have argued that in the CP-violating world the effective field theory cut-off is bounded according to (\ref{4}), and thus quantum stability of the electroweak scale can be attributed to the smallness of the CP violation. This result can not be understood from the effective field theory point of view and requires knowledge of a fundamental theory. Also, CP violation is one of the main ingredients for generating matter -- anti-matter asymmetry in the Universe. Certainly, it would be interesting to deeper understand the origin of the bound discussed in this paper and to study its other implications (e.g., in theory of baryogenesis). 
%\paragraph{Some other implications}
\subparagraph{Acknowledgments.} 
I would like to thank Gia Dvali and Tony Gherghetta for interesting comments and useful suggestions. The work was supported by the Australian Research Council.

%\baselineskip=12pt

%\newpage


\begin{thebibliography}{99}

%\cite{ArkaniHamed:2006dz}
\bibitem{ArkaniHamed:2006dz}
  N.~Arkani-Hamed, L.~Motl, A.~Nicolis and C.~Vafa,
  ``The string landscape, black holes and gravity as the weakest force,''
  JHEP {\bf 0706} (2007) 060
  [arXiv:hep-th/0601001].
  %%CITATION = JHEPA,0706,060;%%

%\cite{Dvali:2007hz}
\bibitem{Dvali:2007hz}
  G.~Dvali,
  ``Black Holes and Large N Species Solution to the Hierarchy Problem,''
  arXiv:0706.2050 [hep-th].
  %%CITATION = ARXIV:0706.2050;%%
  
  %\cite{Dvali:2007wp}
\bibitem{Dvali:2007wp}
  G.~Dvali and M.~Redi,
  ``Black Hole Bound on the Number of Species and Quantum Gravity at LHC,''
  Phys.\ Rev.\  D {\bf 77} (2008) 045027
  [arXiv:0710.4344 [hep-th]].
  %%CITATION = PHRVA,D77,045027;%%
  
  %\cite{Dvali:2008tq}
\bibitem{Dvali:2008tq}
  G.~Dvali, M.~Redi, S.~Sibiryakov and A.~Vainshtein,
  ``Gravity Cutoff in Theories with Large Discrete Symmetries,''
  arXiv:0804.0769 [hep-th].
  %%CITATION = ARXIV:0804.0769;%%

%\cite{Susskind:1995da}
\bibitem{Susskind:1995da}
  L.~Susskind,
  ``Trouble For Remnants,''
  arXiv:hep-th/9501106.
  %%CITATION = HEP-TH/9501106;%%


%\cite{Witten:1979ey}
\bibitem{Witten:1979ey}
  E.~Witten,
  ``Dyons Of Charge E Theta/2 Pi,''
  Phys.\ Lett.\  B {\bf 86} (1979) 283.
  %%CITATION = PHLTA,B86,283;%%

%\cite{Kobakhidze:2008em}
\bibitem{Kobakhidze:2008em}
  A.~Kobakhidze and B.~H.~J.~McKellar,
  ``(De)quantization of black hole charges,''
  arXiv:0803.3680 [hep-th].
  %%CITATION = ARXIV:0803.3680;%%
  
  %\cite{Vafa:2005ui}
\bibitem{Vafa:2005ui}
  C.~Vafa,
  ``The string landscape and the swampland,''
  arXiv:hep-th/0509212.
  %%CITATION = HEP-TH/0509212;%%
  
  
%\cite{Dvali:2006az}
\bibitem{Dvali:2006az}
  G.~Dvali,
  ``Black holes with quantum massive spin-2 hair,''
  Phys.\ Rev.\  D {\bf 74} (2006) 044013 [arXiv:hep-th/0605295]; 
  %%CITATION = PHRVA,D74,044013;%%
 ``Black holes with flavors of quantum hair?,''
  arXiv:hep-th/0607144.
  


\end{thebibliography}
\end{document}